\documentclass[11pt,conference]{IEEEtran}
\usepackage[utf8]{inputenc}
\usepackage{color}
\usepackage{graphicx}
\usepackage{multirow,bigdelim}
\usepackage{booktabs}
\usepackage{caption}% http://ctan.org/pkg/caption
\title{Full-duplex in 5G Small Cell Access: System Design and Performance Aspects}
\title{Full-duplex in 5G Small Cell Access: System Design and Performance Aspects}
\author{Jingwen Bai, Shu-ping Yeh, Feng Xue, Yang-seok Choi, Ping Wang and Shilpa Talwar\thanks{J. Bai, S. Yeh, F. Xue, Y. Choi and S. Talwar are with Intel Labs, Santa Clara, CA, USA; P. Wang was with Intel, now with Apple, USA }}

\usepackage[numbers]{natbib}
\usepackage{graphicx}

\IEEEoverridecommandlockouts
\begin{document}

\maketitle

\thispagestyle{plain}
\pagestyle{plain}

\begin{abstract}
Recent achievement in self-interference cancellation algorithms enables potential application of full-duplex (FD) in 5G radio access systems. The exponential growth of data traffic in 5G can be supported by having more spectrum and higher spectral efficiency. FD communication promises to double the spectral efficiency by enabling simultaneous uplink and downlink transmissions in the same frequency band. Yet for cellular access network with FD base stations (BS) serving multiple users~(UE), additional BS-to-BS and UE-to-UE interferences due to FD operation could diminish the performance gain if not tackled properly. In this article, we address the practical system design aspects to exploit FD gain at network scale. We propose efficient reference signal design, low-overhead channel state information feedback and signalling mechanisms to enable FD operation, and develop low-complexity power control and scheduling algorithms to effectively mitigate new interference introduced by FD operation.
We extensively evaluate FD network-wide performance in various deployment scenarios and traffic environment with detailed LTE PHY/MAC modelling. We demonstrate that FD can  achieve not only appreciable throughput gains~($1.9\times$), but also significant transmission latency reduction~($5$-$8\times$) compared with the half-duplex system.
    
\end{abstract}

\section{Introduction}
5G is expected to provide much higher capacity and quality-of-service beyond what the current 4G network can provide. Revolutionary technologies, including full duplex (FD), are expected to be integrated with existing and evolving systems, e.g. LTE-A and WiFi, to enhance the performance of current systems and meet the new requirements for 5G networks. FD technology supporting the same time-frequency resource transmission and reception can potentially double the spectrum efficiency over the half-duplex~(HD) counterparts. Yet FD operation has long been considered a daunting engineering challenge due to the stringent requirement on self-interference cancellation (SIC). For instance, with a typical transmit power of $21$~dBm for small-cell or mobile station, the self-echo needs to be attenuated by at least 100~dB~\citep{ashu14},

Over the past a few years, with the development in circuit design and advanced signal processing techniques, many researchers have demonstrated the feasibility of FD for short range communication ~\citep{ashu14,Everett14,Sahai11,Bharadia13}, which is within the typically coverage range of small cell access. Based on the recent SIC achievement, one potential application of FD is for small cell access, where FD base stations~(BS) can serve multiple uplink~(UL) and downlink~(DL) users~(UE) in the same time-frequency slot, as shown in Fig.~\ref{fig1}.

{The increased simultaneous active links in FD cellular system, however, could offset the FD gain at network scale due to increased interference. As shown in Fig.~\ref{fig1}, there will be new BS-to-BS and UE-to-UE interference in the FD system in addition to the conventional UL and DL interference that exists in today's HD system. Several papers have studied methods to deal with complex interference introduced by FD} operation~\citep{Goyal15,Han14,Bai16,duplo}. 
%\citep{Goyal15} assumes a powerful genie-aided system where global information including channel state information~(CSI) and transmit power level is perfectly known at all BSs, and studies a network centralized solution to jointly optimize power allocation and UE scheduling decisions. \citep{Han14,Bai16} study the application of large antenna arrays in FD system to mange new interference via beamforming, and no performance evaluation is given in \citep{Han14}. \citep{duplo} is mostly focused on single-cell scenario where  
However, most of the existing works to some extend assume a genie-aided system requiring instantaneous channel feedback, where global channel state information~(CSI) is perfectly known at BSs to enable network cooperation. Moreover, the FD performance is evaluated through Shannon equation. While these works may provide some theoretical bounds on the FD system level performance, it is not clear how much FD gains can actually be achieved in a realistic cellular system as many practical system design aspects are ignored in literature. In this article, our objective is to bridge the 
gaps for practical and efficient interference mitigation schemes which address BS-to-BS and UE-to-UE interference and best exploit full-duplex gain for small cell access. The main contributions of this article are summarized below: 
\begin{itemize}
  \item We propose low-complexity interference mitigation methods to address new interference introduced in FD small cell systems.
  \item We design practical LTE-based signalling mechanisms to enable FD operation, including efficient reference signal design and low-overhead channel state information~(CSI) feedback mechanisms.
  \item We extensively evaluate FD small cell performance in various deployment scenarios and traffic environment with detailed LTE PHY/MAC modelling. We demonstrate that FD can not only achieve upto $1.9\times$ average throughout gain, but also significantly reduce transmission latency with bursty traffic by average 
  $5$-$8\times$. 

\end{itemize}

The remainder of the article is organized as follows. 
We first discuss system challenge and practical consideration for FD small cell access. Next we describe FD system design aspects including reference signal design, CSI feedback mechanisms and interference mitigation methods. Then we present FD performance evaluation results. Finally we conclude the article.

\section{System Challenge and Practical consideration}

\subsection{New Interference in Full-Duplex Cellular Systems}
\subsubsection{Self-Interference}
The echo of the transmitted signal is generally much stronger than the received signal power, thus significantly degrading signal reception if the echo is not suppressed and cancelled below the noise floor. State-of-art self-interference cancellation schemes can achieve more than 110~dB echo attenuation~\citep{Bharadia13} which is sufficient for FD operation in small cell access with low transmit power~\citep{ashu14}. However, for macro cell access with high transmit power, self-interference remains to be an open problem for FD operation.
\begin{figure*}
  \centering
  {\includegraphics[trim=143 100 123 160,clip,width=0.8\textwidth]{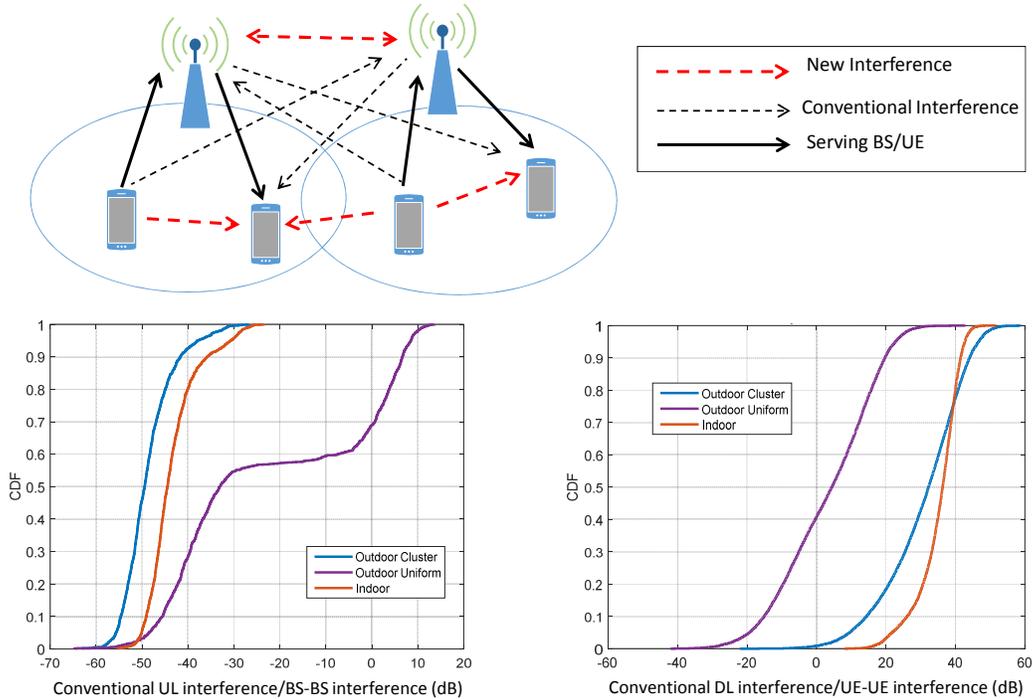}}
  \caption{Complex interference environment in FD cellular system}\label{fig1}
\end{figure*}
%\lipsum[3-10]

\subsubsection{BS-to-BS Interference}
In FD cellular system, DL transmission from neighboring BSs may greatly impact UL reception of the serving BS as shown in Fig.~\ref{fig1}. BS-to-BS channel is more of line-of-sight with much smaller path loss compared to UE-BS channel. Moreover, BS has higher transmit power and antenna gain than UEs. Hence the BS-to-BS interference can easily dominate over desired UL signal if not properly addressed. In Fig.~\ref{fig1}, we compare the strength between conventional UL interference and BS-BS interference in various deployment scenarios. We have considered three typical small cell deployment cases shown in~Fig.~\ref{deploy}: 1) Indoor Hotzone~\citep{889}, where 4 small cell BSs are equally spaced inside a building; 2) Outdoor cluster~\citep{889}, where 4 small cell BSs are dropped inside a cluster in each Macro-cell sector; 3) Outdoor uniform~\citep{828}, where 4 small cell BSs are dropped uniformly in each Macro-cell sector. Each Macro cell site has 3 sectors and operates in a different band. 
All of our deployment and channel models follow 3GPP specification~\citep{889,828} with tier-1~(7 hexagonal cell sites) wrap-around model, and each BS is associated with 10 UL and 10 DL UEs. The conventional UL interference is computed by averaging the interference caused by all other UL UEs to the serving BS. The standard fractional open-loop power control~(OLPC) is also applied to determine the UE transmit power which has been used in the LTE systems~\citep{LTE}. 

From the cumulative distribution function~(CDF) of the ratio between the conventional UL interference and BS-BS interference in Fig.~\ref{fig1}, we can see that BS-BS interference on average is 40-50~dB stronger than the conventional UL interference. Therefore, without mitigating the BS-BS interference, it is almost impossible to transmit data in the UL for  a FD system. 

While BS-BS interference is very dominant, the BS-BS interference channel is rather static since BSs have fixed location, and usually have steady traffic as mobile traffic is more downlink centric. Therefore, static interference mitigation schemes can be applied for BS-to-BS interference mitigation, e.g., elevation nulling~\citep{Yang-seok13} and semi-static power control algorithms which will be described in the next section. 
%\begin{figure}
%  \centering
%  {\includegraphics[trim=150 190 150 200,clip,width=0.7\textwidth]{Figure2.pdf}}
%  \caption{Conventional UL interference versus BS-BS interference for various deployment scenarios.}\label{ULIot}
%\end{figure}

\subsubsection{UE-to-UE Interference}
In the FD cellular system, UL data transmission might interfere with DL data reception, especially when UL UEs and DL UEs are close. Such UE-to-UE interference introduced by FD operation is quite dynamic due to scheduling, UE mobility and non-persistent UL traffic. In Fig.~\ref{fig1}, we show the comparison between conventional DL interference and UE-UE interference. The UE-UE interference is calculated by averaging the interference caused by all other UL UEs to a DL UE. The UE transmit power is determined by fractional OLPC~\citep{LTE}. We can see that for indoor and outdoor cluster cases, the conventional DL interference on average is much stronger than UE-UE interference after OLPC, where each UE has low transmission power. However, as we mentioned earlier, because of the very strong BS-BS interference, additional power boosting is required at UEs to improve UL performance in the FD system. Hence the downlink performance will be degraded afer UL power boosting due to stronger UE-UE interference. 

The UE-UE interference includes both intra-cell UE-UE interference which is caused by the same-cell UL UEs as well as inter-cell UE-UE interference which is caused by neighboring-cell UL UEs. 
Intra-cell UE-UE interference can be mitigated via smart scheduling to prevent strong interfering UL-DL UE pair from transmitting at the same resource. And inter-cell UE-UE interference can be further mitigated through joint scheduling with cross-cell coordination. We will describe our proposed FD scheduling algorithms in the next section.
%\begin{figure}
%  \centering
%  {\includegraphics[trim=150 190 150 200,clip,width=0.7\textwidth]{Figure3.pdf}}
%  \caption{Conventional DL interference versus UE-UE interference for various deployment scenarios.}\label{DLIot}
%\end{figure}

\subsection{Practical System Consideration}
Most past works are focused on deriving the optimal solutions for best FD capacity gain by ignoring many practical implementation aspects. The most related work to our problem setup is~\citep{Goyal15}, where a network centralized solution is studied to mitigate new interference in FD cellular system. \citep{Goyal15} requires perfect global information~(such as CSI and power level) to jointly optimize power allocation and scheduling decisions with BS cooperation. While tight coordination offers better performance in theory, in reality it requires high signaling overhead which could offset the benefits from BS cooperation.
Moreover, the FD performance in the existing literature~\citep{Yang-seok13,Goyal15,Bai16} is evaluated by calculating rates using Shannon equation, where only signal-to-interference-and-noise ratio~(SINR) is approximated under large scale fading. 
While such analysis and evaluation methodology provides theoretical bounds on FD network performance, they fail to fully capture practical system behaviors nor to reveal insight into practical FD system-level performance.

%\textcolor{red}{Timing issue?}
In this article, we design low-complexity interference mitigation methods without cross-cell coordination. We evaluate FD small cell performance with detailed LTE Physical Layer~(PHY) and Medium Access Control Layer~(MAC) modelling, where system throughput is calculated based on LTE MAC layer transport block size. 
We demonstrate that our proposed approach is sufficient in most small cell deployment scenarios to achieve significant throughput gain for a realistic FD small cell system. 

\section{Full-Duplex System Design Aspects}
\subsection{Reference Signal Design}
Reference signals in the existing HD LTE networks including Cell-specific Reference Signal~(CRS), Demodulation Reference Signal (DMRS), Sounding Reference Signal~(SRS), etc., are used to estimate channel for feedback and decoding. There are some general criterion for the reference signal design in the FD system. Firstly, we need to protect existing reference signals for channel estimation, such as CRS. Further, we suggest symmetric design for uplink and downlink transmission to facilitate interference management and FD operation. We propose to reserve certain resources to measure the overall UE-UE interference caused by all the scheduled UL UEs by muting the DL transmission in the corresponding resources. Hence the DL UEs can compute the SINR based on the measured UE-UE interference and existing CRS~(which measures conventional DL interference) for both feedback and decoding. 

We also suggest a new UE-UE interference measurement reference signal~(UE-UE IM-RS) structure to enable joint-scheduler by measuring UE-UE interference for each UL-DL UE pair. BS will configure UL UEs to send a set of (quasi-) orthogonal UE-UE IM-RS sequences. BS will inform the DL UE of the UL UE index group so that the DL UE can measure the interference power from each UL UE in the group. Without cross-cell coordination, only the intra-cell UE-UE interference can be measured for each UL-DL UE pair. 

The actual position of such UE-UE IM-RS in the frame structure depends on the individual timing advance (TA) of the each UL-DL UE pair as well as the propagation delay between UL and DL UE in the pair. Such suggested reference signal design can be quite efficient as it only reserves a few additional resources.

\subsection{Channel State Information Feedback Mechanisms}
In the current LTE system, each DL UE needs to compress measured SINR from the reference signals, map the SINR to a Channel Quality Indicator~(CQI) value which indicates the supported modulation and coding scheme~(MCS) by allowing certain block error rate~(BLER)~(e.g., 10\%), and finally report such CQI value to the BS. In a MIMO system, the UE also feedback RI and PMI values, known as Rank Indicator and Precoding Matrix Indicator, respectively, for MIMO spatial multiplexing.

The existing CQI report in LTE system does not take into account the UE-UE interference introduced by the FD operation. Hence we propose new CSI feedback mechanisms to track the new UE-UE interference from FD operation. 
Based on the above-mentioned FD reference signal designs, new CQI for each UL-DL UE pair based on measured UE-UE interference can be reported to the BS. In order to perform joint-scheduling, the BS needs to gather such CQI report for all the serving UL-DL UE pairs in each sub-band~(usually consists of several resource blocks for scheduling), which will incur huge feedback overhead.

Therefore it is of critical importance to reduce the feedback overhead. We then propose low-overhead wide-band feedback mechanisms which requires low update frequency: 1) 1-bit feedback: A bitmap of 0 or 1 to indicate whether a particular UL-DL UE pair can be scheduled in the same resources or not; 
2) Multi-bit feedback: finer granularity to describe the CQI degradation for each UL-DL UE pair.
These proposed wide-band feedback mechanisms can be used together with conventional sub-band CQI feedback~(in the LTE system) to enable a smart joint scheduler for FD operation.

\subsection{Elevation Beamforming and Power Control}
As we discussed earlier, the BS-BS interference is very dominant. Hence we perform a two-step approach for BS-BS interference mitigation. First, BSs perform elevation beamforming to create a null at vicinity of 90 degree~\citep{Yang-seok13}. As demonstrated in~\citep{Yang-seok13}, upto 35 dB nulling can be achieved at transmitter and receiver allowing 2 degree variation. With higher antenna height variation, less BS-BS interference suppression can be obtained. 
Note that, for enterprise deployment with BS deployed on the ceiling and antenna facing downwards, the antenna beam-pattern is typically designed with 20-40dB attenuation in 90$^{\circ}$ elevation. Therefore, even without additional design for FD-BS, substantial amount of BS-to-BS interference suppression can be achieved.  

Next we deal with the residual BS-BS interference by applying uplink power control. Our proposed power control algorithm is based on the standard fractional OLPC~\citep{LTE}. In OLPC, the UE transmission power per resource block is determined by $P_{tx}=\min\{P_{max},~P_0+\alpha\times\mathrm{PL}\}$, where $P_{max}$ is the maximum UE power, $P_0$ is a semi-static base level, $\alpha$ is the path-loss compensation factor and $\rm PL$ is the path-loss component. 
We can control UE power by adjusting the semi-static base level $P_0$. There are two criterion to determine the parameter $P_0$ for power control: 1) The modified transmission power should be large enough to overcome BS-BS interference; 2) The increased UE power should not exceed certain threshold to prevent UE-UE interference from degrading DL performance. Based on long-term channel statistics, the UE power boosting factor can be decided.

\subsection{Scheduling Algorithms}
First we introduce the FD basic scheduler, where 
DL and UL UEs are scheduled independently based on their CSI feedback which incorporates UE-UE interference measurement based on our reference signal design. Without loss of generality, we adopt proportional-fairness~(PF) scheduling. In the FD basic scheduler, the PF metric is computed individually for DL and UL based on conventional feedback mechanisms. Hence there is no explicit protection for DL transmission nor additional feedback overhead. Next, we describe our joint scheduler, where the objective is to maximize the joint UL and DL utility function. Based on collected per UL-DL UE pair feedback as explained in the previous section, joint scheduler can select the best UL-DL UE pair for transmission which achieves the highest joint PF metric. Then BS will perform rate matching and precoder selection~(in MIMO) for DL and UL separately. The joint scheduler can protect DL transmission by mitigating intra-cell UE-UE interference. The associated overhead for joint scheduler depends on the corresponding feedback mechanisms proposed earlier.   

Our designed FD schedulers do not require cross-cell coordination, thus only mitigating intra-cell interference.
For scenarios with strong inter-cell interference, we suggest to use semi-static frequency planning approaches such as fractional frequency reuse which can achieve good complexity-performance tradeoff.

\section{Performance Evaluation}
\subsection{Full-duplex System-Level Simulator}
In order to evaluate FD small cell performance, we have developed a FD system-level simulator~(SLS). In our SLS, we model the complicated interference environment under FD operation, where spacial channel model~(SCM) is used to generate fast fading channels~\citep{814}. Light-of-sight~(LoS), non-LoS path-loss and shadowing models follow 3GPP specifications for different small cell deployment scenarios~\citep{889,828}. 
Our SLS has detailed LTE PHY/MAC modelling, including Minimum Mean Square Error-Interference Rejection Combining~(MMSE-IRC) receiver, non-ideal delayed CSI feedback, Mean Mutual Information per coded Bit (MMIB) mapping for PHY abstraction, Hybrid Automatic Repeat Request~(HARQ), OLPC, link adaption with MCS selection targeting 10\% BLER, Transport Block~(TB) size mapping, etc.

\begin{figure}
  \centering
  {\includegraphics[trim=100 110 60 100,clip,width=0.68\textwidth]{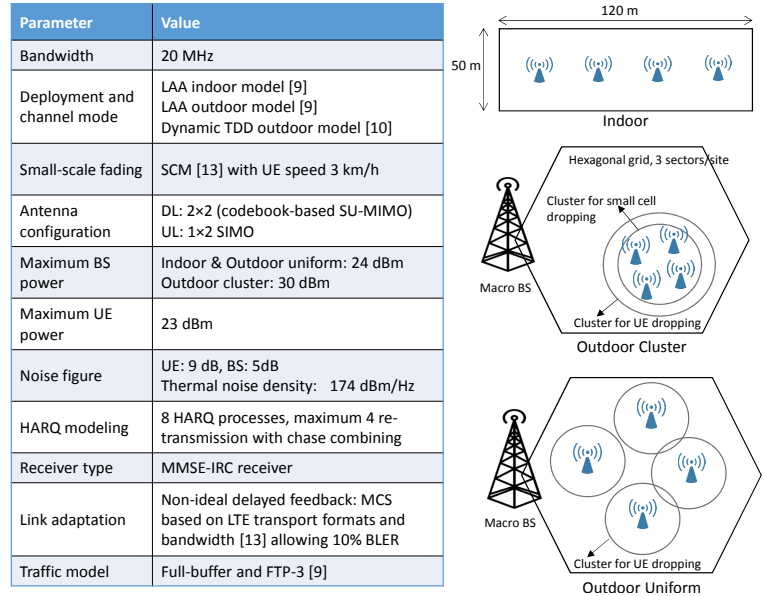}}
  \caption{Different small-cell deployment scenarios and simulation assumptions.}\label{deploy}
\end{figure}

We consider three deployment scenarios: indoor, outdoor cluster and outdoor uniform~(see Fig.~\ref{deploy}) for FD performance evaluation with tier-1~(7 hexagonal cell sites) wrap-around model, and each BS is associated with 10 UL and 10 DL UEs. We compare 20 MHz FD system with 10 MHz frequency-division duplex~(FDD) UL and 10 MHz FDD DL system. We assume symmetric operation, i.e., OFDMA in both DL and UL. Some key simulation assumptions are given in the table in Fig.~\ref{deploy}, and the details can be found in ~\citep{889,828,814}. We demonstrate the FD performance assuming FD BSs with half-duplex UEs\footnote{Similar FD performances are observed with full-duplex UEs but with reduced feedback overhead.}. 
We run several random drops with sufficiently long simulation time for both FD and FDD systems, and compute the respective throughput based on LTE MAC layer TB size~\citep{LTE}.
% \begin{table}
% \begin{center}
% \caption{Simulation Assumptions}  \label{table1}
% \vspace{-2mm}
% \def\arraystretch{1.2}%  1 is the default, change whatever you need}
%     \begin{tabular}{ | l | p{5cm} |}
%     \hline
%     \textbf{Parameter} & \textbf{Value}\vspace{1mm} \\\hline
%     Bandwidth& 20 MHz\\ \hline
%     Deployment and & LAA indoor model~\citep{889} & channel model & LAA outdoor model~\citep{889} & & Dynamic TDD outdoor model~\citep{828} \\\hline
%     Small-scale fading  &  SCM~\citep{814} with UE speed 3 km/h \\ \hline
%     Antenna configuration & DL: 2$\times$2 (codebook-based SU-MIMO) UL: 1$\times$2 SIMO \\ \hline
%     Maximum BS power& Indoor \& Outdoor uniform: 24 dBm Outdoor cluster: 30 dBm\\\hline
%     Maximum UE power & 23 dBm \\ \hline
%     Noise figure&UE: 9 dB, BS: 5dB && Thermal noise density: $-$174 dBm/Hz \\ \hline
%     HARQ modeling& 8 HARQ processes, maximum 4 retransmission with chase combining\\ \hline
%     Receiver type& MMSE-IRC receiver \\ \hline
%     Link adaptation & Non-ideal delayed feedback: MCS based on LTE transport formats and bandwidth~\citep{814} allowing 10\% BLER \\ \hline
% %scheduler type & Proportional-fairness scheduler\\\hline
%     Traffic model& Full-buffer and FTP-3~\citep{889}\\ \hline
%     \end{tabular}
%     \end{center}
% \end{table}

\subsection{Full Buffer Traffic}
We start with the case of full buffer traffic to evaluate our proposed interference mitigation algorithms, where all interferences are present. The FD gain defined as the throughput ratio between FD and FDD is given in Table~\ref{table2} when different interference mitigation methods is applied. 
To allow BS antenna height variations under actual deployment, we assume BS has 20~dB elevation nulling at transmitter and receiver, respectively. Thus total 40 dB BS-BS interference suppression can be achieved. From Table~\ref{table2}, we can see that in dense deployments like indoor and outdoor cluster environments, UL performance can be severely degraded by BS-BS interference. Even after applying 20 dB elevation nulling at BS, the BS-to-BS interference is still so dominant that no UL FD gain can be observed. With our purposed UL power boosting technique as described previously, we can recover UL FD gain to above 1.9$\times$. For sparse deployment like outdoor uniform environment, 40 dB BS-BS nulling can suppress most BS-BS interference.

While more than 1.8$\times$ average UL FD gain can be achieved with BS-nulling and power control, the increased UE power will degrade DL performance due to stronger UE-UE interference.
In particular, for outdoor uniform which represents sparse deployment, interference from UEs within the same cell significantly limit DL FD gain. With our proposed joint scheduler, intra-cell UE-UE interference can be effectively mitigated to obtain 1.83$\times$. For dense deployment, the joint scheduler also recovers DL FD gain caused by UL power boosting to 1.9x and 1.81x for indoor and outdoor cluster deployment, respectively. 
Such performance achieved by joint scheduler requires full per UL-DL UE pair sub-band feedback which will incur significant feedback overhead. With less than $2\%$ additional feedback overhead (e.g., 4-bit pair-wise wide-band feedback with low update frequency), we show that 98\% of joint-scheduler performance can be retained.
\hspace{-5mm}\begin{table}[htbp]
  \centering
  \caption{Average full-duplex gain with different interference mitigation methods}
  \label{table2}
  \vspace{-2mm}
    \begin{tabular}{|p{1cm}|p{0.4cm}|p{1.7cm}|p{1.9cm}|p{1.9cm}|}
    \toprule
    \multicolumn{2}{|c|}{\textbf{Deployment }} & \multicolumn{3}{|c|}{\textbf{Interference Mitigation Schemes}} \\
\cmidrule{3-5}    \multicolumn{2}{|c|}{\textbf{Scenarios}} & Basic scheduler~\& & Basic scheduler~\&  & Joint scheduler~\&  \\
    \multicolumn{2}{|l|}{} &  BS-nulling &  BS-nulling~\& &  BS-nulling~\& \\
    \multicolumn{2}{|l|}{} &       & Power control & Power control \\
    \midrule
    Indoor & DL    & 2.0x  & 1.78x & 1.9x \\
\cmidrule{2-5}          & UL    & 0.95x & 1.94x & 1.96x \\
    \midrule
    Outdoor & DL    & 2.0x  & 1.7x  & 1.81x \\
\cmidrule{2-5}    cluster & UL    & 0.6x  & 1.91x & 1.92x \\
    \midrule
    Outdoor  & DL    & 1.51x & 1.31x & 1.78x \\
\cmidrule{2-5}    uniform & UL    & 1.71x & 1.83x & 1.84x \\
    \bottomrule
    \end{tabular}%
\end{table}%

\subsection{Non-Full Buffer Traffic}
The real-world traffic is often bursty and sporadic. Moreover, there is usually more traffic in DL than in UL. As a result, 3GPP is already considering flexible UL and DL resource allocation such as
dynamic time-division duplex~(TDD) for unpaired spectrum ~\citep{828} and flexible-duplex~\citep{ran-86} for paired spectrum.

The drawback of dynamic TDD is the complicated timing design for HARQ which could increase feedback latency. And flexible duplex will face adjacent band interference and require additional guard band. In contrast, FD system will impose no complication on HARQ design, reduce guard period and guard band in addition to flexible traffic adaptation with sufficient self-echo cancellation. 

We next show the performance comparison of 20 MHz FD, 10 MHz FDD UL + 10 MHz FDD DL, and 20 MHz flexible duplex. We assume an ideal flexible duplex system where each sub-band can be scheduled for \emph{either} UL \emph{or} DL, and no guard-band, no adjacent-band interference, no additional overhead is considered. We further assume total 40 dB BS-nulling to suppress BS-BS interference in flexible duplex (same as FD system).

The performances of all three systems are tested under realistic bursty traffic using FTP-3 traffic model~\citep{889} with PF scheduling, where the file arrival process for each UE follows Poisson process. 
We assume file size is 0.1 Mbyte, with DL:UL traffic ratio = 2:1. We use perceived throughput as the performance metric. The perceived throughput during active time is defined as the size of a burst divided by the time difference between reception of the last packet of a burst and 
arrival of the first packet of a burst. 
\begin{figure}
  \centering
  {\includegraphics[trim= 200 110 250 125,clip,width=0.65\textwidth]{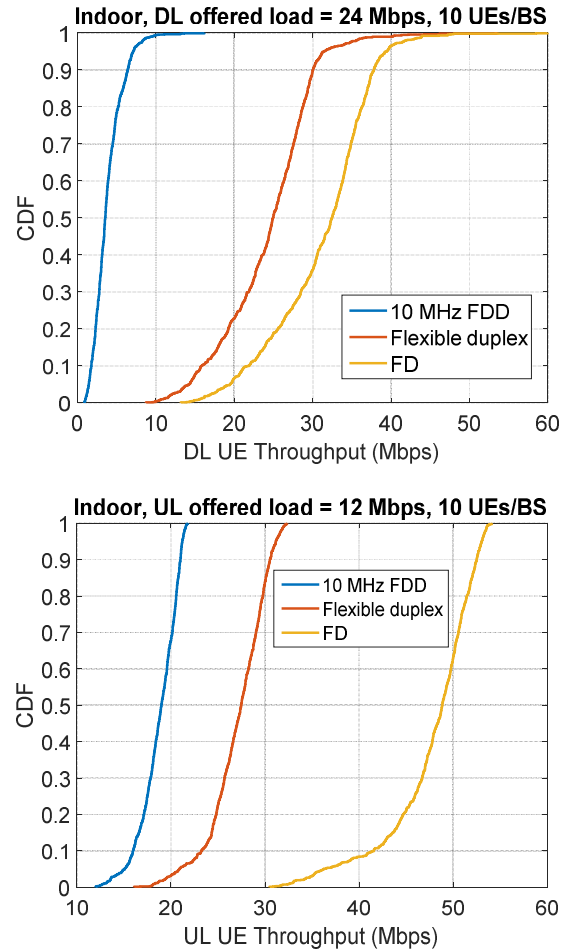}}
  \caption{DL and UL perceived throughput CDF of FDD, flexible duplex and FD systems with bursty traffic.}\label{Dltput}
\end{figure}
%\begin{figure}
%  \centering
%  {\includegraphics[trim= 180 190 150 160,clip,width=0.64\textwidth]{NFBULTputCDF.pdf}}
 % \caption{UL perceived throughput CDF of FDD, flexible duplex and FD systems with bursty traffic.}\label{Ultput}
%\end{figure}

\begin{figure*}
  \centering
  {\includegraphics[trim= 20 160 40 155,clip,width=0.75\textwidth]{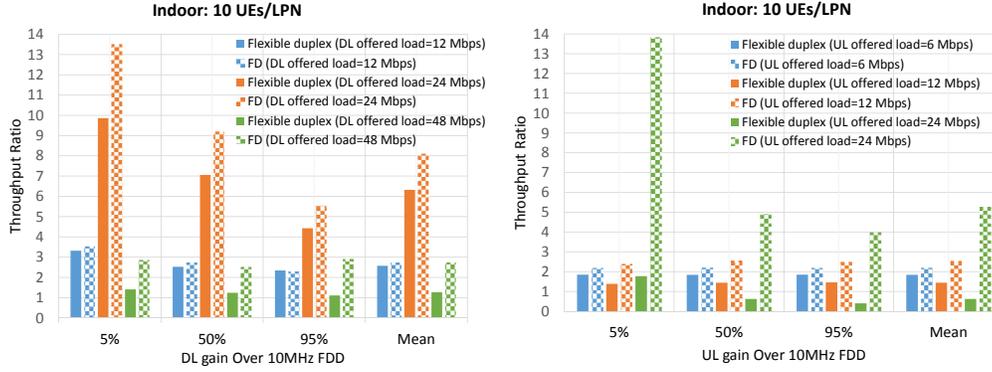}}
  \caption{DL and UL perceived throughput ratio compared with FDD under various traffic loads. }\label{Dltputgain}
\end{figure*}

Fig.~\ref{Dltput} depicts the CDFs of DL and UL perceived throughput for FDD, flexible duplex and FD, under 24 Mbps DL offered load with 12 Mbps UL offered load. We can see that significant improvement can be achieved by FD over FDD in DL and over flexible duplex in UL. Note that the FD performance is based on basic scheduler.

The DL and UL perceived throughput gains over FDD of 5-percentile, 50-percentile, 95-percentile and average UEs for flexible duplex and FD systems under various loads are given in  Fig.~\ref{Dltputgain}, respectively. By varying the offered traffic load from low to medium to high in DL, and very low to low to medium in UL~(due to 2:1 DL-UL traffic ratio), we conclude that on average, FD can achieve about $2\times$ under low load due to double spectrum. While under medium load, $3$-$8\times$ FD gain in DL and UL can be obtained due to reduced queueing delay in addition to double spectrum. Under high load, FD gain will reduced to $2\times$ due to increased interference.   
In contrast, flexible duplex fails to provide gain over FDD as load increases, and is always inferior to FD.

Fig.~\ref{Dltputgain} demonstrates the significant 5-percentile FD gain in both UL and DL. Because the cell-edge UEs have lower rate, they will experience longer queueing delay. FD system can significantly reduce the transmission latency, thus yielding upto $14\times$ FD gain for cell-edge UEs in both UL and DL.

%\begin{figure}
 % \centering
 % {\includegraphics[trim= 200 190 150 190,clip,width=0.72\textwidth]{NFBULTputGain.pdf}}
 % \caption{UL perceived throughput ratio compared with FDD under various traffic loads.}\label{Ultputgain}
%\end{figure}

\section{Conclusions}
In this article, we consider practical FD system design aspects for small cell access. Efficient reference signal design, CSI feedback mechanisms are proposed to enable low-complexity interference mitigation methods which prevent BS-BS and UE-UE interference from diminishing FD gains. With extensive performance evaluation under various deployment scenarios and traffic environment with our LTE-based system level simulator, we conclude that not only practical FD gain can be retained at network scale and but also  significant transmission latency reduction can be achieved from FD operation. 

\bibliographystyle{ieeetr}  
\footnotesize
\bibliography{references}
\end{document}